\documentclass[12pt]{article}
\usepackage{enumerate}
\usepackage{amsmath}
\usepackage{hyperref} 
\hypersetup{colorlinks, breaklinks, urlcolor=blue, linkcolor=blue} 

\usepackage[utf8]{inputenc}

\usepackage{graphicx}
\usepackage{float}
\usepackage{wasysym}
\usepackage{authblk}

\numberwithin{equation}{section}

\begin{document}

\title{Gamma ray signals of the annihilation of Higgs-portal singlet dark matter}
\author{Frederick S. Sage and Rainer Dick}
\affil{Department of Physics and Engineering Physics, University of Saskatchewan,\\116 Science Place, Saskatoon, SK S7N 5E2, Canada}

\date{}

\maketitle

\abstract{This article is an exploration of gamma ray signals of annihilating Higgs-portal singlet scalar and vector dark matter. Gamma ray signals are considered in the context of contributions from annihilations of singlets in the galactic halo to the Isotropic Gamma Ray Background (IGRB), in the context of the Galactic center excess, and in the context of observations of dwarf spheroidal galaxies. We find that Higgs-portal singlets of both species with a mass of $~$65 GeV can explain the Galactic center excess with reasonable accuracy, but that this mass range is in tension with current direct detection bounds. We also find that singlets in the mass range of 250-1000 GeV are consistent with both the Fermi-LAT IGRB observations and direct detection bounds. Additionally, bounds from gamma ray observations of the dwarf spheroidal galaxy Segue I are translated into bounds on the Higgs-portal couplings.}

\vspace{2cm}

One of the most intriguing issues that exists in both particle physics and astrophysics research is the dark matter problem. Gravitational evidence has been accumulating for some time that the majority of matter in the universe is nonluminous and nonbaryonic, and the lack of any suitable candidates in the Standard Model of particle physics has led to vast amounts of literature exploring theories of particle dark matter beyond the Standard Model.  To account for cosmological evidence, particle dark matter is frequently assumed to satisfy the Weakly Interacting Massive Particle (WIMP) paradigm, in which the current dark matter abundance can be related to the WIMP annihilation cross sections at the time in the early universe at which the WIMP is decoupled from radiation. The literature surrounding particle dark matter is vast; several excellent reviews exist \cite{Bergstrom:2012fi}\cite{Bertone:2010zza}.

Dark matter is usually assumed to have negligible electromagnetic interactions and very weak interactions of other types with Standard Model particles, leading to only a few possible detection mechanisms. The basic methods of searching for particle dark matter include nuclear recoil searches, collider searches, and indirect searches for annihilation products. Nuclear recoil, or direct, searches are usually considered to be the primary search mechanism, and provide the strongest exclusion limits on the properties of particle dark matter. Modern collider searches are hadronic and are very complicated, making extraction of any kind of positive dark matter signal exceedingly difficult without knowing what energy range to look at. Indirect searches look for the annihilation products of dark matter particles in the galactic halo. Such products must be both stable and detectable to provide any useful signal, and this limits potential indirect signals to be photons, stable antimatter (positrons, antiprotons and light antinuclei) or neutrinos. Of these, the high energy photon signal is considered the most promising for a variety of reasons. Despite large astrophysical backgrounds which mask potential signals, gamma rays are relatively simple to detect. In addition, gamma rays propagate more or less unimpeded through the galaxy, allowing searches to focus on specific regions, which is not possible with other searches.

This article is a discussion of the gamma ray indirect search prospects of Higgs-portal singlet dark matter. With the next generation of direct detection experiments expected to report results within a few years, minimal Higgs-portal models are coming very close to either full exclusion or observation. We find it an appropriate time to revisit the potential for indirect detection of these models, focusing on the gamma ray channel, which has seen a recent increase in interest. 

In section 1 we provide an overview of the models that we explore in this article. We describe their properties and discuss where they have previously appeared in the literature. In section 2, we apply several constraints to the parameter space of the models to reduce the mass range that needs to be considered. We describe the gamma ray spectrum that is the result of the annihilation of Higgs-portal singlets in section 3. Section 4 contains a discussion of Higgs-portal singlets in the mass range of 63 GeV to 70 GeV and whether or not they can explain the observed Galactic center gamma ray excess that was recently reported. In section 5, we discuss a higher mass region of a few hundred GeV that is in better agreement with direct detection constraints and is more consistent with the diffuse gamma flux observed by the Fermi gamma ray telescope and make comparisons with bounds derived from the observations of Earth-based Cherenkov gamma ray telescopes. In section 6 we summarize our conclusions.

 \section{Higgs-portal singlets}

Minimal models of scalar-portal hidden sectors, frequently interacting with baryonic matter through couplings to the Standard Model Higgs boson, have been explored in great depth over the last decade. It is as important to explore extensions to the Standard Model motivated by minimality as it is to explore those motivated by symmetry considerations. Despite their relatively simple structure, these models remain phenomenologically interesting, both in their own right and as low-energy approximations to more complicated theories. In many ways, minimal Higgs-portal singlets are examples of archetypal WIMPs.

We work with two models containing fields that are singlets under the gauge transformations of the Standard Model of particle physics, interacting at leading order only through gravitational interactions and with the Standard Model Higgs doublet $H$. The first model has a scalar singlet $S$, with a Lagrangian density of

\begin{equation}
\mathcal{L}_S =\mathcal{L}_{SM}  -\frac{1}{2}\partial^\mu S \partial_\mu S -\frac{1}{2}m_{S}^2S^2-\frac{\lambda_S}{4}S^4-\frac{\eta_S}{2}S^2 H^{\dagger}H.\end{equation}

The second model has a vector singlet $V^\mu$ with a Lagrangian density of

\begin{equation}
\mathcal{L}_V =\mathcal{L}_{SM}-\frac{1}{4}V^{\mu\nu}V_{\mu\nu}-\frac{1}{2}m_{V}^2V_\mu V^\mu
-\frac{\lambda_V}{4}\left(V_\mu V^\mu \right)^2- \frac{\eta_V}{2}V_\mu V^\mu H^{\dagger}H.\end{equation} 

We have introduced a field strength tensor $V^{\mu\nu}=\partial^\mu V^\nu - \partial^\nu V^\mu$ for the massive Higgs-portal vector singlet. In unitary gauge, which makes the phenomonological content of the models explicit, the Lagrangian densities take the forms:

\begin{equation}
\mathcal{L}_S =\mathcal{L}_{SM}  -\frac{1}{2}\partial^\mu S \partial_\mu S -\frac{1}{2}m_{S}^2S^2-\frac{\lambda_S}{4}S^4-\frac{\eta_S v_h}{2}S^2 h -\frac{\eta_S}{4}S^2h^2,\end{equation}

\begin{equation}
\mathcal{L}_V =\mathcal{L}_{SM}-\frac{1}{4}V^{\mu\nu}V_{\mu\nu}-\frac{1}{2}m_{V}^2V_\mu V^\mu 
-\frac{\lambda_V}{4}\left(V_\mu V^\mu \right)^2-\frac{\eta_V v_h}{2}V_\mu V^\mu h - \frac{\eta_V}{4}V_\mu V^\mu h^2.\end{equation} 

In the above, the Standard Model Higgs boson field is $h$, and $v_h$ is the vacuum expectation value of the Higgs field, which we take to be the standard value of 246 GeV \cite{Agashe:2014kda}. The parameters $\eta_S$ and $\eta_V$ are the Higgs-portal couplings, which regulate the strength of the interaction between the singlet fields and the Higgs boson. The parameters $\lambda_S$ and $\lambda_V$ govern the singlet self-interactions. 

The bare masses of the singlet fields are $m_{S0}$ and $m_{V0}$. We remain agnostic as to by what mechanism these masses are generated, noting that these mass terms as they are violate no symmetry of the Lagrangian. There has been investigation into dynamical generation of these mass terms \cite{Steele:2013fka}. In the following analysis, we will work with the physical masses of the singlets as they appear in unitary gauge, which take into account contributions from the Higgs-portal interactions. In terms of the bare masses, the physical masses are $m_S^2=m_{S0}^2+\eta_S v_h^2/2$ and $m_V^2=m_{V0}^2+\eta_V v_h^2/2$.

These models are minimal extensions of the Standard Model. A $Z_2$ symmetry has been imposed to ensure absolute stability of the singlets. This also has the effect of preventing any mixing between the scalar singlet and the Higgs, which would complicate the phenomenology. It also excludes any interaction terms from the Lagrangian other than those already present. These models are generic enough that they can be considered as effective degrees of freedom of more complicated higher energy theories. The phenomenology at the low energy scale of hundreds of GeV remains similar. The vector singlet model also hints at the presence of a dark gauge sector of some kind, but we refrain from exploring specific models \cite{Lebedev:2011iq}\cite{Baek:2012se}. 

There is also the possibility of a fermionic Higgs-portal singlet. The dynamics of such a field are described by a Lagrangian of the form
\begin{equation}
\mathcal{L}_\chi = \bar{\chi}\left( i\gamma^\mu \partial_\mu - m_{\chi 0}\right)\chi - \frac{1}{\mu_f}\bar{\chi}\chi H^{\dagger}H \end{equation}
\begin{equation}
=
\bar{\chi}\left( i\gamma^\mu \partial_\mu - m_{\chi}\right)\chi - \frac{v_h}{\mu_f}\bar{\chi}\chi h - \frac{1}{2\mu_f}\bar{\chi}\chi h^2 \end{equation}

where the second line is in unitary gauge.

This model is not minimal, in the sense that a new mass scale $\mu_f$ is required, which explicitly includes additional dynamical content. In the next section, we note that the constraints from the requirement that the singlet abundance match the observed cold dark matter abundance require this mass scale to be in the hundreds of GeV range. Because of the existence of this extra parameter at such a low energy scale, we exclude fermionic Higgs-portal singlets from this study.

The scalar singlet model was originally proposed in the context of dark matter by Silviera and Zee \cite{zee1985}, and later examined in more detail by McDonald \cite{mcdonald1994}. After these initial explorations, the model was left untouched until direct detection experiments began construction, when several authors returned to the model \cite{burgess2001}\cite{oconnell2007}\cite{davoudiasl2005}\cite{holz2001}\cite{mcdonald2002}\cite{wilczek2006}  . Since then, the model has appeared regularly in the literature, with analysis being done on direct detection \cite{yaguna2009}\cite{he2009}\cite{Han:2015hda}, indirect detection through gamma rays \cite{Dick2008}\cite{yaguna2009}\cite{Okada:2013bna}, indirect detection through antimatter \cite{yaguna2009b} and production at particle colliders \cite{he2010a}\cite{barger2008}\cite{farina2011}\cite{tanden2011}\cite{Han:2016gyy}. Much of the phenomenology of the scalar model has been reviewed and updated recently \cite{PhysRevD.88.055025}. The model is also of interest due to the effects it can have of the stability of the electroweak vacuum and electroweak symmetry breaking \cite{gonderinger2010}\cite{Steele:2013fka}.

\section{Parameter space constraints}

Each of the above models has two phenomenologically relevant parameters, the mass of the singlet and the Higgs-portal coupling. The singlet self-interaction strength is not phenomenologically accessible and so we ignore it. We require only that it is small enough to keep the theory in the perturbative regime. 

Making the assumption that one species of singlet accounts for all of the observed dark matter, and that singlets were produced thermally following the standard scenario for cold dark matter freeze-out, we can reduce the dimension of the parameter space. 

The abundance of a particle species $n(t)$ can be found by solving the rate equation; when thermal production is possible, as is the case for dark matter in the early universe, a thermal term is included, gives

\begin{equation}
\frac{dn}{dt} + 3nH(t)=-\langle\sigma v\rangle\left( n^2-n_{eq}^2\right).\end{equation}

The equilibrium number density $n_{eq}$ can be approximated analytically as in the standard Lee-Weinberg theory \cite{weinberg1977}. Through a series of manipulations, this equation can be cast into a form relating the mass of the particle and its freeze-out temperature. It has been shown that this expression can be approximated for thermal dark matter \cite{Sage2015} by the simpler form

\begin{equation}
m[\mathrm{GeV}]=\varrho_{CDM}\left[\frac{\mathrm{eV}}{\mathrm{cm}^3}\right] 2.333\times 10^{-11}\exp\left(m/T_f\right).\end{equation}

By finding the solution of the rate equation at the present time and inserting the observed dark matter abundance, we can obtain a constraint on the thermally averaged annihilation cross section $\langle \sigma v \rangle_{Th}$:

\begin{equation}
\langle \sigma v \rangle_{Th} \left( T_f\right) = \frac{2m-3T_f}{2m-T_f}\frac{2.075m \times 10^{-24} \mathrm{cm}^3/\mathrm{s}}{\varrho_{CDM}\left[\mathrm{eV}/\mathrm{cm}^3\right] T_f}. \end{equation}

Noting that (2.3) is an expression for the cross section required to produce the observed abundance that does not depend explicitly on the Higgs-portal coupling $\eta$, we compare to the thermally averaged cross sections computed semianalytically using standard methods. The relativistic cross sections appear in the appendix, and we perform the velocity weighted thermal averaging by following Gondolo and Gelmini \cite{PhysRevD.56.1879}\cite{Gondolo1991145}, who have derived the expression
\begin{equation} 
\langle \sigma v \rangle \left( T_f\right) = \frac{1}{8m^4 T_f K_2^2 \left( m/T_f \right)} \int_{4m^2}^\infty ds \sigma(s) K_1\left( \frac{\sqrt{s}}{T_f}\right) \sqrt{s} \left( s- 4m^2\right). \end{equation}

In this expression $m$ is the mass of the annihilating particle and $K_1 \left( x\right)$ and $K_2 \left( x\right)$ are the modified Bessel functions of the first and second order respectively. The fully relativistic cross section as a function of the center of mass energy $s$ is $\sigma \left( s\right)$.

By extracting the $\eta$ dependence, we can obtain a formula for $\eta$ that is required by the dark matter balance equation (2.3):

\begin{equation}
\eta^2 = \frac{\langle \sigma v \rangle_{Th} }{\langle \sigma v \rangle / \eta^2}. \end{equation}

This is an expression that detrmines the $\eta$ for which singlets satisfy the observed cold dark matter abundance in the Lee-Weinberg approximation as a function of singlet mass. Application of this constraint reduces the parameter space of the theory by a dimension from $(m,\eta)$ to $(m)$. 

We have used the updated 2015 Planck results \cite{Ade:2015xua} for the dark matter abundance. The thermally constrained $\eta^2$ values for scalar and vector singlets are plotted against singlet mass in Figure 1 for the mass region that is of interest to us. These values remain in the perturbative regime for singlet masses well over 10 TeV.

\begin{figure}
\begin{center}\includegraphics[scale=0.4]{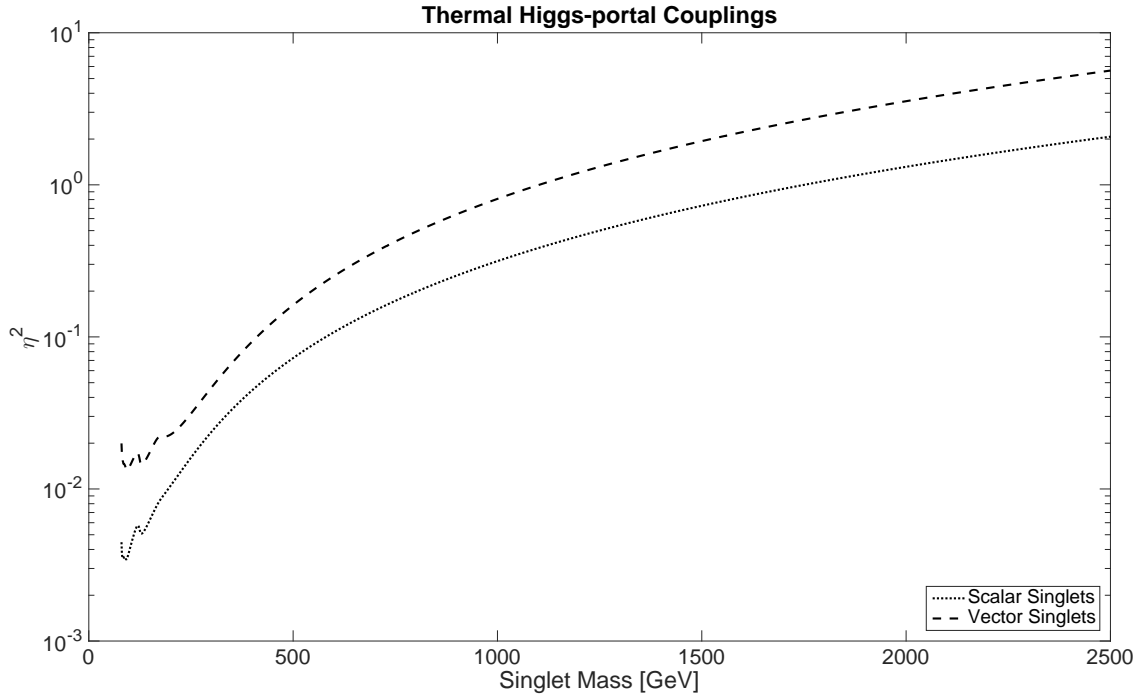}\end{center}
\caption{Thermally constrained couplings $\eta_S^2$ and $\eta_V^2$ as obtained from (2.5) for the dark matter abundance reported in \cite{Ade:2015xua}. The bumps at the low end of the mass range are due to the effects of the Higgs resonance in the annihilation cross section.}
\end{figure}

For the fermionic Higgs-portal singlet case, the analysis proceeds in the same fashion. The annihilation cross sections are of course different, and the parameter $\mu_f^2$ is constrained rather than $\eta_{S,V}$. We have found that $m_\chi =200$ GeV requires $\mu \approx 900$ GeV and decreases as $m_\chi$ increases.

Caution should be taken in the region immediately below 62.5 GeV, where the Higgs decay channel to singlets opens, as our approximations break down. This is due to strong resonance effects in the annihilation cross sections disrupting the relation as presented.

The next constraint we apply to restrict our mass range is due to the stringent limits on the invisible Higgs boson decay width. If the singlet is less than half the Higgs decay width, an invisible decay path for the Higgs into two singlets is opened through the Higgs-portal, with decay rates of 

\begin{equation}
\Gamma_{h\rightarrow SS}=\frac{\eta_S^2 v_h^2}{32\pi m_h^2}\sqrt{m_h^2-4m_S^2},\end{equation}

\begin{equation}
\Gamma_{h\rightarrow VV}=\frac{\eta_V^2 v_h^2}{64\pi m_h^2}\sqrt{m_h^2-4m_V^2}\frac{\left(m_h^2-2m_V^2\right)^2+8m_V^4}{m_V^4}\end{equation}

for the scalar and vector species, respectively. The mass $m_h$ is the mass of the Standard Model Higgs, taken to be 125.9 GeV \cite{Agashe:2014kda}. The vacuum expectation value of the Higgs field is as before $v_h=246$ GeV. We compute and display the invisible Higgs branching ratio, 

\begin{equation}
BR_{inv}=\frac{\Gamma_{inv}}{\Gamma_{inv}+\Gamma_{vis}}\end{equation}

in Figure 2. Because our thermal constraint on the coupling becomes intractable just below half the Higgs mass, we plot the branching ratios for a representative set of Higgs-portal couplings which are consistent with extrapolation through the resonance region. The Figure also includes the bounds on the invisible Higgs decay width as reported by ATLAS \cite{PhysRevLett.112.201802} and CMS \cite{Chatrchyan:2014tja}. It is clear that the bounds on the invisible Higgs decay width are too stringent to allow thermally constrained Higgs-portal singlets in the mass range $m_S,m_V<m_h/2$. We ignore that mass region in our analysis because of these bounds, and the resonance-induced breakdown of the approximation.

\begin{figure}
\begin{center}\includegraphics[scale=0.4]{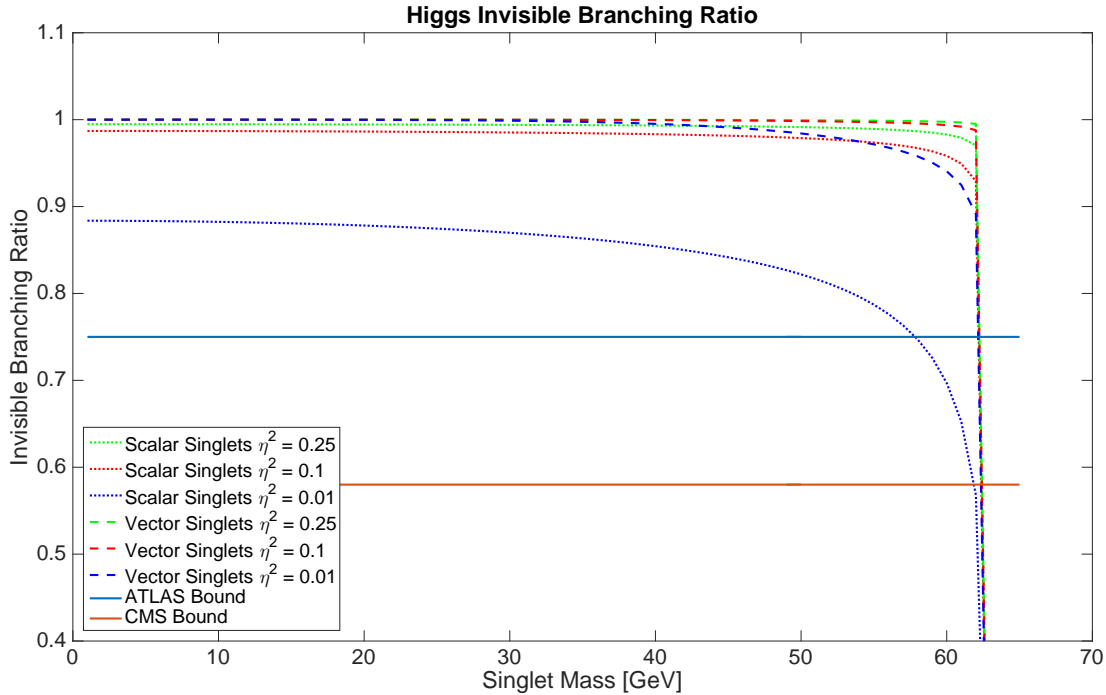}\end{center}
\caption{Higgs invisible decay branching ratio (2.8) for $m_S,m_V<$65 GeV compared with bounds from ATLAS \cite{PhysRevLett.112.201802} and CMS \cite{Chatrchyan:2014tja}. Values of $\eta^2$ chosen to illustrate the dependence of the width on the coupling.}
\end{figure}

Other collider based bounds on the properties of dark matter, such as missing energy searches \cite{Djouadi201265}, are not strong enough to restrict the parameter space of the models. 

Direct detection is the primary means of searching the dark matter parameter space, and the bounds it provides are strong. Current generation direct detection experiments \cite{Aprile:2012vw}\cite{PhysRevLett.109.181301}\cite{Faham:2014hza} constrain the WIMP-nucleon recoil cross section to be less than around $10^{-44.5}\:\mathrm{cm}^2$ for most mass values in the GeV to thousands of GeV range. It has been shown that next generation direct detection experiments such as XENON1T \cite{Aprile:2015uzo} and DEAP3600 \cite{Amaudruz:2014nsa} will fully test the parameter space of the scalar and vector Higgs-portal models \cite{Sage2015} in the mass range above 200 GeV. Even current generation experiments rule out certain low mass regions of the parameter space. With the thermal constraint on the couplings discussed above (2.5), we can use reported exclusion bounds on nuclear recoil cross sections to constrain the mass range in which the model remains consistent with these results. 

The nuclear recoil cross section for scalar and vector Higgs-portal singlets is \cite{Sage2015} 

\begin{equation}
\sigma_{NR} =\frac{g_{hNN}^2\eta_{S,V}^2 v_h^2}{4\pi m_h^4}\frac{m_N^2}{\left(m_{S,V}+m_N\right)^2}.\end{equation}

The singlet and Higgs couplings and masses are as labelled previously. We take the nucleon mass $m_N$ to be 930.5 MeV, which is approximately the average value of the mass of a nucleon in stable liquid xenon or argon isotopes. The parameter $g_{hNN}$ describes the interaction strength between the Higgs boson and nucleons. Despite the additional helicity states in the massive vector case, the cross sections for both the scalar and vector species reduce to the same expression (2.9) in the nonrelativistic limit. 

The Higgs-nucleon coupling was originally calculated several decades ago \cite{shifman1978} and has been of limited interest \cite{cheng1988} until the advent of nuclear recoil direct detection experiments relatively recently. This calculation has also been revisited recently and updated with modern lattice and experimental results \cite{Cheng:2012qr}. 

This quantity depends on the strangeness content of the nucleon

\begin{equation}
y_N = \frac{2\langle N \left| \bar{s}s \right| N \rangle}{\langle N \left| \bar{u}u+\bar{d}d \right| N \rangle} \end{equation}

which is a notoriously poorly constrained parameter. Originally, this quantity was assumed to be very small, but more recent studies indicate this is not the case \cite{PhysRevD.85.054502}\cite{PhysRevD.85.054510}\cite{PhysRevD.87.114510}
\cite{PhysRevD.88.014503}\cite{PhysRevD.91.094503}\cite{PhysRevD.88.054507}
\cite{PhysRevC.89.045202}\cite{Stahov:2012ca}\cite{Alarcon:2012nr}. We work with the conservative range of values as discussed in \cite{Sage2015}:

\begin{equation}
210 \: \mathrm{MeV} \le g_{hNN}v_h \le 365 \: \mathrm{MeV}.\end{equation}

The nuclear recoil cross sections for both species computed using this range of values for the Higgs-nucleon coupling are presented in Figure 3, compared to reported bounds from nuclear recoil experiments \cite{Aprile:2012vw}\cite{PhysRevLett.109.181301}\cite{Faham:2014hza}. A basic visual analysis can find that the XENON100 reported results rule out scalar singlets with thermally constrained couplings with a mass below 70 GeV and vector singlets with thermally constrained couplings with a mass below 130 GeV. The LUX constraints are stronger, ruling out scalars with masses below 105 GeV and vectors with masses below 185 GeV. We will revisit the implications of these constraints in section 4.

\begin{figure}
\begin{center}\includegraphics[scale=0.4]{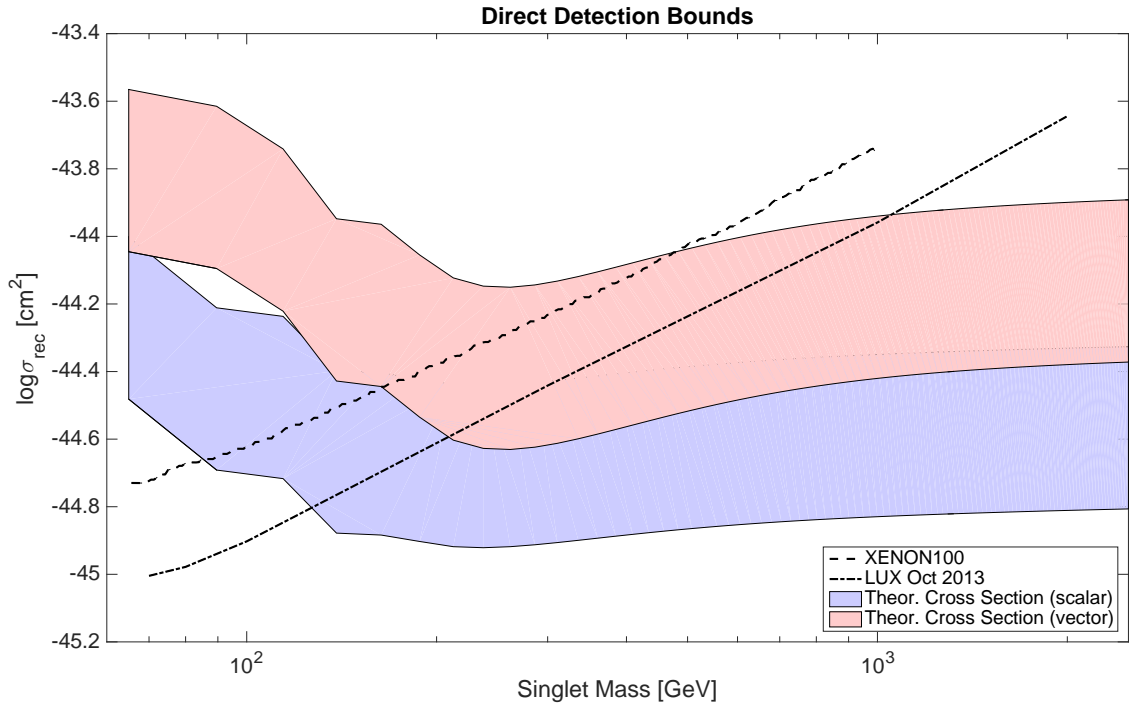}\end{center}
\caption{Current nuclear recoil exclusion bounds compared to the theoretical nuclear recoil cross section for the scalar and vector Higgs-portal singlet models. The shaded regions cover the variation in the cross section introduced by the uncertainty in the Higgs-nucleon coupling parameter $g_{hNN}$.}
\end{figure}

\section{Gamma rays from singlet annihilations}

Dark matter in the galactic halo can annihilate, and annihilation to Standard Model particles will lead to the production of high energy gamma rays. These gamma rays can be detected by gamma ray telescopes on Earth and in Earth orbit, allowing for the possibility of an observable signal of annihilating dark matter. This section discusses what form a potential annihilation signal will take when observed by these instruments. Much of this material is generic, and is independent of the particle physics model used for the dark matter.

The differential gamma ray flux per energy and solid angle from annihilating dark matter is given by \cite{Bergstrom1998137} 
\begin{equation}
\frac{d\Phi_\gamma}{d\Omega dE}=\frac{1}{2}\frac{r_{\astrosun}}{4\pi}\left(\frac{\rho_{\astrosun}}{m_{DM}}\right)^2\times J\times \left(\sum_f \langle \sigma v\rangle_f \frac{d\mathcal{N}^f_\gamma}{dE}\right).\end{equation}

An additional factor of $1/2$ is required when the dark matter particles are not self-conjugate. The energy spectrum of photons produced by annihilation of dark matter into a final state $f$ is $d\mathcal{N}^f_\gamma /dE$. We discuss this function and how it is obtained further below.

The factor $J$ represents an integral over the dark matter distribution along a line of sight. It describes how many annihilation events contribute to the differential flux. It is given by  

\begin{equation}
J = \int_{l.o.s}\frac{ds}{r_{\astrosun}}\left(\frac{\rho\left(r\left(s,\theta\right)\right)}{\rho_{\astrosun}}\right)^2.\end{equation}

The solar position $r_{\astrosun}$ and local dark matter density $\rho_{\astrosun}$ are conventional weight factors. The radial coordinate is related to the line of sight $s$ and the aperture angle $\theta$ between the line of sight and a line connecting the telescope to the Galactic Center by

\begin{equation}
r\left(s,\theta\right) = \sqrt{r_{\astrosun}^2+s^2-2r_{\astrosun} s \cos\theta}.\end{equation}

These expressions can be related to the galactic polar coordinates $(d,b,l)$ which are defined by $x=d\cos(b)\cos(l)$, $y=d\cos(b)\sin(l)$ and $z=d\sin(b)$. Our solar system is at $x=y=z=0$ so that the Galactic center is at $x=r_{sun}$, $y=z=0$. The angle $\theta$ can be expressed as $\cos (\theta)=x/d=\cos(b)\cos(l)$.

When discussing the flux from a specific region of interest as observed by an instrument, a more useful quantity is the flux averaged over that region $\Delta\Omega$. To compute this quantity, it is conventional to write it in terms of an averaged $J$ factor, denoted $\bar{J}$. This averaged factor is given by

\begin{equation}
\bar{J}\left( \Delta\Omega\right) = \frac{1}{\Delta\Omega}\left( \int_{\Delta\Omega} J d\Omega \right).\end{equation} 

For a specific region of interest $\Delta\Omega$, the differential flux from that region is then

\begin{equation}
\frac{d\Phi_\gamma}{dE}=\frac{1}{2}\frac{r_{\astrosun}}{4\pi}\left(\frac{\rho_{\astrosun}}{m_{DM}}\right)^2\times \bar{J}\Delta\Omega\times \left(\sum_f \langle \sigma v\rangle_f \frac{d\mathcal{N}^f_\gamma}{dE}\right).\end{equation}

More details on the calculation of these quantities is available in the literature \cite{Cirelli:2010xx}.

In calculation of these $\bar{J}$ factors for different observational regions of interest, we use the parameters used in the article whose results we are comparing against. The default values are $\rho_{\astrosun}=0.3 \:\mathrm{GeV}/\mathrm{cm}^3$ \cite{Agashe:2014kda} and $r_{\astrosun}=8.33\:\mathrm{kpc}$ \cite{Gillessen:2008qv}. The quantity $\bar{J}$ remains very sensitive to astrophysical uncertainties, and the vast majority of the uncertainty in our results comes from these astrophysical uncertainties.

In this article, we assume the dark matter distribution in the galaxy follows the spherically symmetric NFW profile \cite{navarro1996}, which remains the standard profile used in particle astrophysics. The NFW profile takes the form

\begin{equation}
\rho \left( r \right) = \rho_s \left(\frac{r_s}{r}\right) \left(1+\frac{r}{r_s}\right)^{-2}\end{equation}

where $r_s$ and $\rho_s$ are scale parameters of the profile. 

The photon spectrum $d\mathcal{N}^f_\gamma /dE$ resulting from a given final state $f$ is in general a very complicated function to compute to any reasonable degree of accuracy. This is true even when $f$ is restricted to Standard Model states, as is the case with Higgs-portal singlet models. This article uses the PPPC 4 DM ID results \cite{Cirelli:2010xx} for final state photon spectra. These results were produced using the PYTHIA Monte Carlo event generator \cite{Sjostrand2008852}\cite{1126-6708-2006-05-026}, including parton showers and hadronization. Electroweak corrections were also included in these results \cite{Ciafaloni:2010ti}. 

Higgs-portal singlets annihilate to all Standard Model particles that interact with the Higgs field, including the Higgs boson itself. At tree level, this includes all quarks and charged leptons as well as the massive gauge bosons. Annihilation to photons and gluons occurs at the 1-loop level. Annihilation to neutrinos depends on the neutrino mass generation mechanism and is negligible regardless. The annihilation cross sections to fermions are proportional to the masses of those fermions through their Yukawa couplings to the Higgs, leading to differences in magnitude related to the mass hierarchy. With this in mind, we somewhat arbitrarily choose to include only the following fermion annihilation channels: $\bar{c}c$, $\bar{b}b$, $\bar{t}t$, $\tau^+ \tau^-$. Contributions from lighter fermions are suppressed by a factor of at least $10^{-2}$. Annihilation into $W^+W^-$, $ZZ$ and $hh$ final states is also taken into account. It should be noted that some of these channels will be kinematically disallowed at lower singlet masses. For reference, Figure 4 contains plots of the branching ratios to these final states as a function of singlet mass for the thermally constrained couplings discussed in the previous section.

\begin{figure}
\begin{center}\includegraphics[scale=0.4]{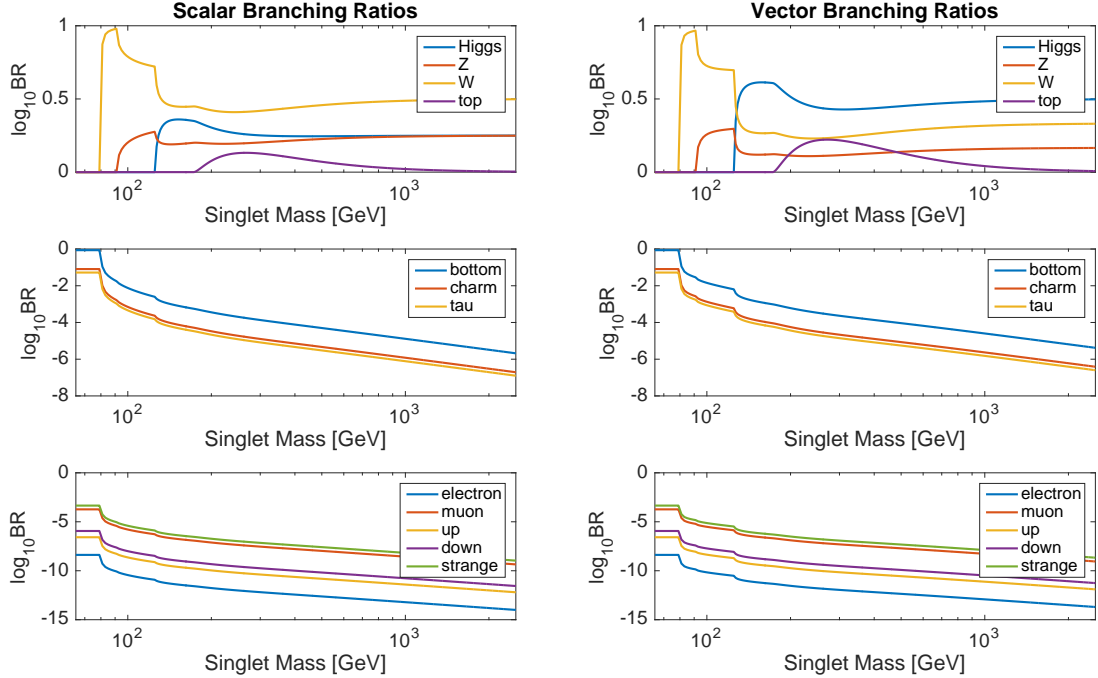}\end{center}
\caption{Branching ratios for the annihilation of scalar and vector Higgs-portal singlets on a double-logarithmic scale.}
\end{figure}

Contributions to the continuum flux from direct annihilations to photons and gluons are suppressed by factors of $1/m_W^2$ and $1/m_t^2$ respectively compared to the tree level annihilations, and so can be safely neglected. These annihilation cross sections are also discussed in the appendix.

Annihilation to Standard Model particles proceeds through Higgs-mediated channels, with the relativistic cross sections appearing to leading order in the appendix. The nonrelativistic limits of the corresponding velocity-weighted cross sections that are required for the calculation of gamma ray fluxes are obtained by multiplication with $2k/\sqrt{k^2+m_{S,V}^2}$ and taking the limit $k\rightarrow 0$, where $k$ is the magnitude of the three-momentum of the initial state in the center of momentum frame. 

Only prompt photons directly produced in the annihilation process are considered in this article. Not included are secondary photons from processes such as final state bremsstrahlung, synchrotron radiation from propagating antimatter final states, and most significantly gamma rays from inverse Compton scattering of antimatter annihilation products off background photons. Since the energy fraction of stable antimatter produced is much lower than that of photons due to the mass scale suppression by the Higgs Yukawa couplings, neglecting these couplings is justified. We also neglect photons from extragalactic annihilations, considering only annihilations of dark matter in the Milky Way halo.

Observations by gamma ray telescopes show the existence of a diffuse flux of high energy photons. It is possible that annihilations of dark matter particles in the galactic halo contribute to this flux, and so studying this flux may lead to constraints on the properties of dark matter. We are interested in possible contributions from Higgs-portal singlet annihilations to the Isotropic Gamma Ray Background (IGRB). We work with the results of the Fermi-LAT collaboration \cite{Ackermann:2014usa}, obtained after 50 months of observations. The IGRB is what remains of the observed diffuse flux after known sources have been subtracted. It includes both galactic and extragalactic components. For some dark matter models, the photon spectrum resulting from annihilations in the halo are similar in magnitude to the IGRB, allowing for meaningful comparisons.

The Fermi-LAT collaboration parameterizes the IGRB spectrum by a power-law with exponential cutoff:

\begin{equation}
\frac{d \mathcal{N}}{dE} = I_{100} \left( \frac{E}{100\:\mathrm{MeV}} \right)^{-\alpha}\exp \left( -E/E_{cut} \right).\end{equation}

The fit parameters are reported for three different foreground models (A, B, C) as discussed in the reference. The differences between the foreground models do not impact our conclusions, although in general, the effects of different foreground and background gamma ray emission models can be very important \cite{Zhou:2014lva}.

We close this section by mentioning that there is another significant means of searching for a WIMP annihilation signal which we have not considered, which is the annihilation line search. This kind of search looks for photons produced by the direct annihilation $XX\rightarrow \gamma\gamma$, which can occur only at higher loop orders for nearly all WIMP dark matter models. The result is a line in the gamma ray spectrum at the mass of the dark matter particle. We do not consider line searches, referring the interested reader instead to several recent publications \cite{Duerr:2015mva}\cite{Duerr:2015aka}\cite{Duerr:2015bea}, which use Fermi data to search for lines from the annihilation of Higgs-portal singlets. While we do not perform the analysis, for completeness we include the relativistic annihilation cross sections $\sigma_{SS\rightarrow \gamma\gamma}$ and $\sigma_{VV\rightarrow\gamma\gamma}$ in the appendix.

\section{Galactic center excess}

As discussed in section 2, assuming that Higgs-portal singlets make up the entirety of observed dark matter and were produced thermally in the early universe allows us to reduce the parameter space to a single dimension, the singlet mass. We now consider certain mass regions motivated by their potential for an observable gamma ray signal. 

Detection of a gamma ray annihilation signal of dark matter has always been a difficult prospect because of the relatively small photon production rate compared to other astrophysical processes. For this reason, the Galactic Core with its expected higher density of dark matter has frequently been proposed as a potential target \cite{Springel:2008by}. Over the last few years, several groups have analyzed the data produced by the Fermi-LAT gamma ray telescope and have reported statistically significant evidence of a gamma ray excess at the GeV scale in the region of the Galactic center \cite{Goodenough:2009gk}\cite{Hooper:2010mq}. This excess has recently been confirmed with a careful analysis by the Fermi-LAT collaboration \cite{TheFermi-LAT:2015kwa}. The excess is spherically distributed in space, which favors an interpretation that sources the excess through the annihilation of a diffuse particle distribution. This interpretation is consistent with WIMP dark matter that follows standard distribution profiles.

The excess has been much explored in the literature \cite{Daylan:2014rsa}\cite{Calore:2014nla}\cite{Agrawal:2014oha}\cite{Calore:2014xka}, often with comparisons to generic WIMP dark matter. These generic discussions tend to focus on annihilation to a $\bar{b}b$ final state, which unfortunately does not include important contributions from gauge and Higgs boson channels. It has also been discussed in the context of specific Higgs-portal dark matter models \cite{Agrawal:2014oha}\cite{Duerr:2015bea}\cite{1475-7516-2014-09-013}\cite{Alvares:2012qv}\cite{Kim:2016csm},  as well as with effective scalar-portal operators \cite{Alves:2014yha}\cite{Beniwal:2015sdl}. Very recently, a detailed statistical analysis of the scalar model in the context of the Galactic center excess has been made available \cite{Cuoco:2016jqt}. 

We use the results reported in \cite{Abazajian:2014fta}, which characterize the excess in a region of interest of $7^{\circ}\times 7^{\circ}$ centered on the Galactic center ($\left| l \right| ,\left| b \right| \le 3.5^{\circ}$) using both a `log-parabola' spectrum 

\begin{equation}
\frac{d \mathcal{N}}{dE} = N_0 \left( \frac{E}{E_b} \right)^{-\left(\alpha +\beta \ln \left( E/E_b \right) \right) }\end{equation}

and a broken power law spectrum

\begin{equation}
\frac{d \mathcal{N}}{dE} = N_0 \left( \frac{E}{E_0} \right)^{-\gamma_c}\exp \left( -E/E_c \right).\end{equation}

The numerical values resulting from their fit, which we use in our calculations, are available in the reference \cite{Abazajian:2014fta}. 

The energy scale of the Galactic center excess indicates that for it to be caused by annihilating dark matter, that dark matter must be in the tens of GeV range. In particular, once the annihilation channels to electroweak bosons open, more high energy photons are produced than appear in the excess. With this in mind, we explore the lowest possible mass region allowed by invisible Higgs decay bounds, 63 GeV to 70 GeV. It turns out that 63 GeV is too close to the $s$-channel resonance from the Higgs mediator, leading to flux predictions nearly an order of magnitude too large, so we exclude anything below 63.5 GeV.

The predicted fluxes from the annihilation of Higgs-portal singlets of masses 64 GeV, 65 GeV, 66 GeV, 67 GeV, 67.5 GeV and 70 GeV are compared to the Galactic center excess as reported in \cite{Abazajian:2014fta}. For both the scalar and vector cases, a 66-67.5 GeV singlet compares reasonably well to the excess, with other masses matching less well. The main discrepancy is that the predicted flux falls off more slowly than the observed excess. 

In Figures 5 and 6 we plot the predicted fluxes for the annihilation of scalar and vector singlets, respectively, in the region of interest around the Galactic center. The mass that matches the observed fits best is the 67 GeV singlet for both species, though the 67.5 GeV singlet is also close. Higher masses produce fluxes that are far too small, as can be seen even at 70 GeV. Decreasing the mass further increases the flux as the singlet mass approaches the half-Higgs mass resonance. The displayed regions for these fits are taken from the reported errors on the fit parameters. Our results agree reasonable well with those of  \cite{Cuoco:2016jqt}. Our values for the scalar mass are somewhat larger than their reported best fit value of 62.7 and further away from the highly constrained region near the resonance that occurs at $m_S=m_h/2$.

\begin{figure}
\begin{center}\includegraphics[scale=0.4]{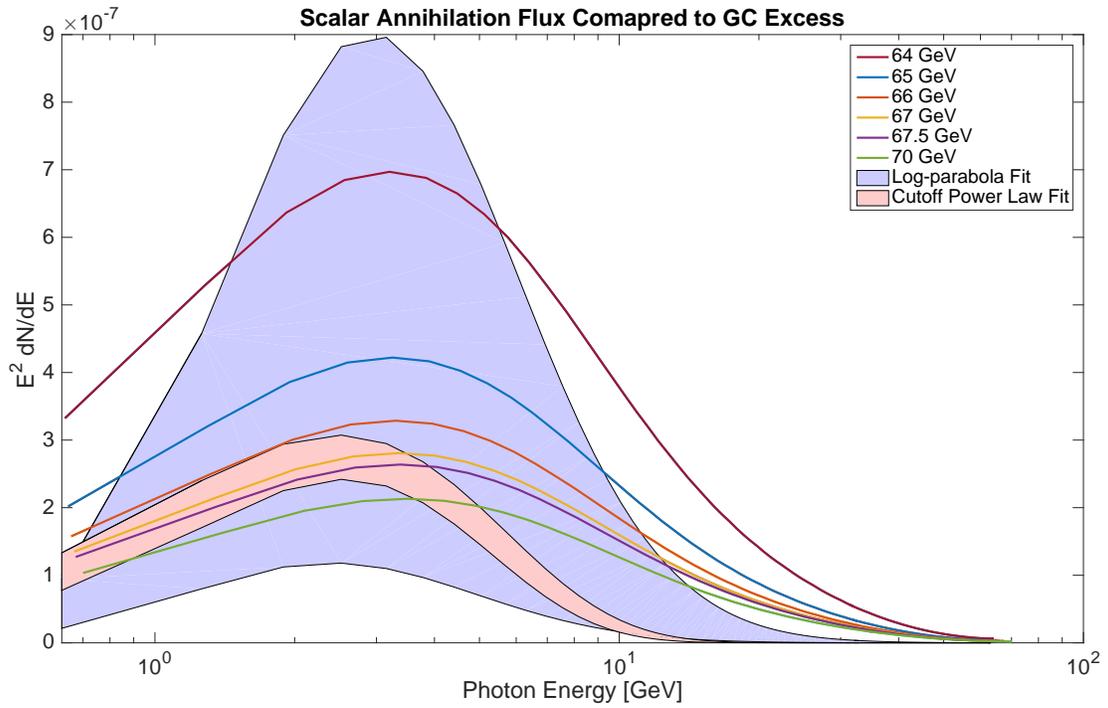}\end{center}
\caption{Flux produced by the annihilation of scalar singlets of masses in the range 64 GeV to 70 GeV in the $7^\circ\times 7^\circ$ region of interest around the center of the Galaxy compared to two fits with error regions as reported in \cite{Abazajian:2014fta}. Fit 1 is the log-parabola spectrum (4.1) and Fit 2 is the broken power law spectrum (4.2).}
\end{figure}

\begin{figure}
\begin{center}\includegraphics[scale=0.4]{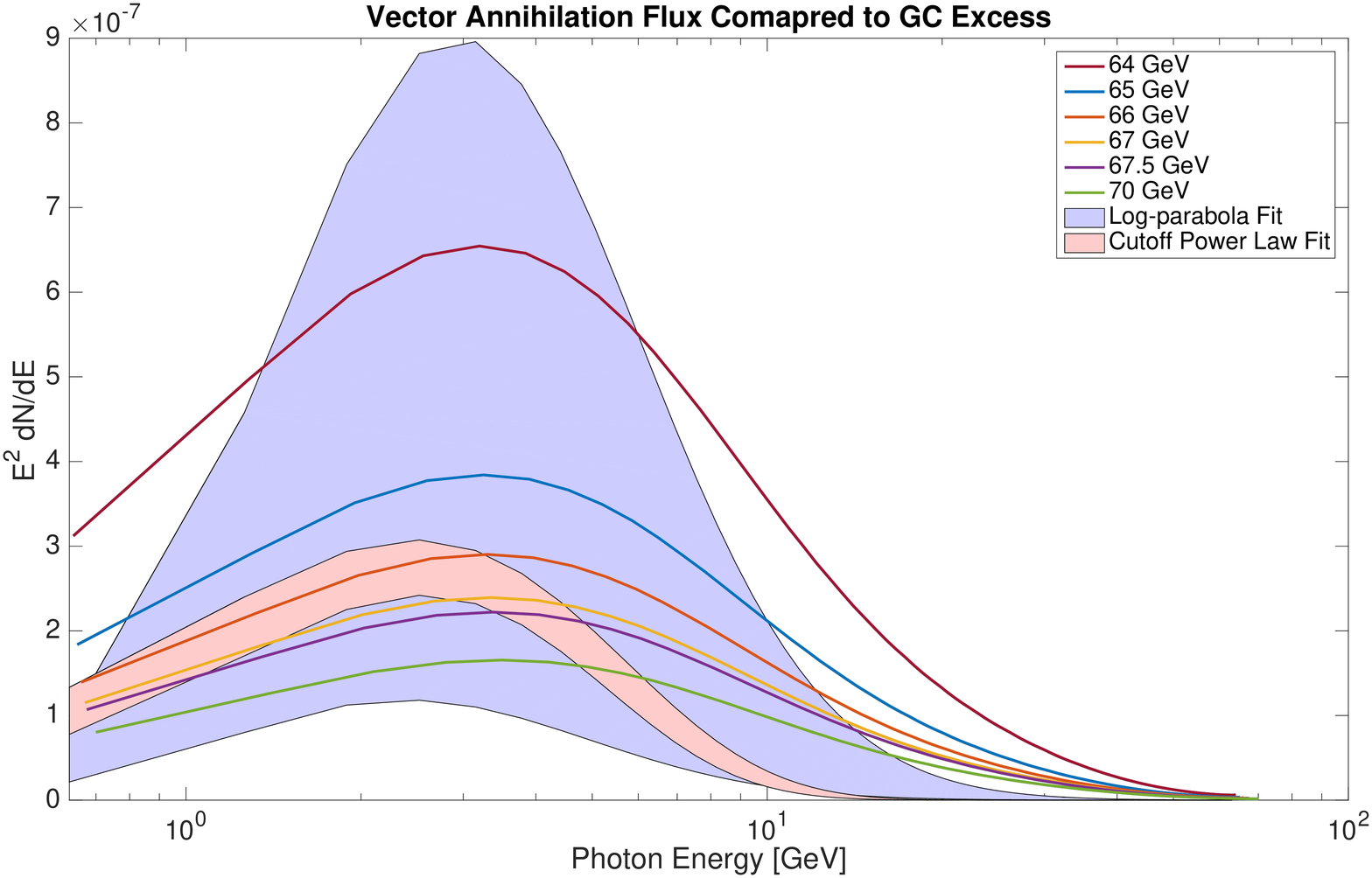}\end{center}
\caption{Flux produced by the annihilation of vector singlets of masses in the range 64 GeV to 70 GeV in the $7^\circ\times 7^\circ$ region of interest around the center of the Galaxy compared to two fits with error regions as reported in \cite{Abazajian:2014fta}. Fit 1 is the log-parabola spectrum (4.1) and Fit 2 is the broken power law spectrum (4.2).}
\end{figure}

The mass region explored here can explain the Galactic center excess to some degree, but is it consistent with the IGRB spectrum as observed by Fermi-LAT? To answer this question, we find the ratio of the predicted flux from annihilating Higgs-portal singlets in the galactic halo and the reported fit to the IGRB spectrum $\left( d\Phi_\gamma^{Fermi} / dE \right) / \left( d\Phi_\gamma^{Theory} / dE \right)$. When this ratio is smaller than one, the predicted flux from annihilating singlets is less than the observed flux and so is consistent with the explanation that some part of the IGRB is produced by singlet annihilations. When this ratio is larger than one, the predicted flux is larger than what is observed, and some tension with the IGRB observations is introduced.

As discussed in section 3, the astrophysical uncertainties that go into the computation of the theoretical flux are quite large, and can have significant effects on the final result. To exhibit their effects on this ratio in the relevant mass range, we present the ratios plotted in Figure 7 for three different values of the cold dark matter density $\rho_{\astrosun}$. The usual assumption in most of the literature is $\rho_{\astrosun} = 0.3\: \mathrm{GeV}/\mathrm{cm}^3$ \cite{Agashe:2014kda}, but other values are occasionally used. Our variation comes from the standard error range of $\pm 0.1 \:\mathrm{GeV}/\mathrm{cm}^3$, though higher values have been reported in the literature \cite{1475-7516-2013-07-016}. As can be seen in Figure 7, varying the dark matter density can have significant effects on the ratio. The ratio presented uses foreground model A, but the other models do not appreciably alter the conclusions.

Even with nonstandard values for the dark matter density, Figure 7 shows that in certain regions of the energy spectrum, the calculated theoretical flux is larger than the observed Fermi IGRB. Though the ratio shows the fluxes remain within an order of magnitude, this comparison indicates there is some tension between the IGRB observations and the interpretation of the Galactic center excess as annihilating Higgs-portal singlets.

\begin{figure}
\begin{center}\includegraphics[scale=0.4]{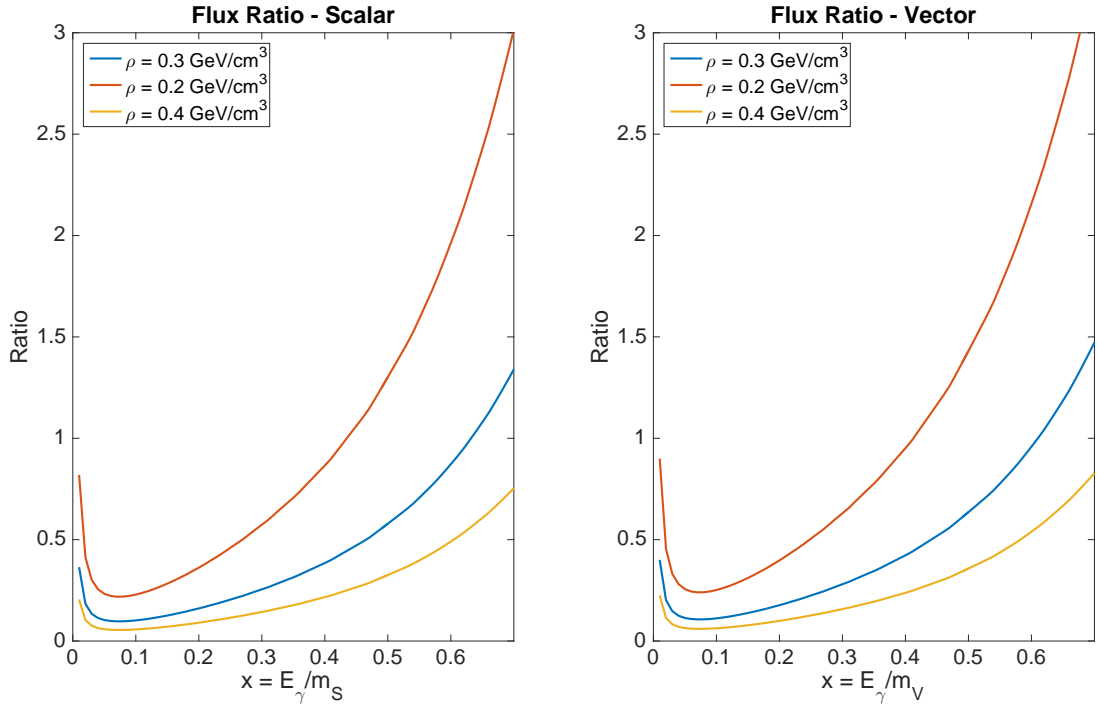}\end{center}
\caption{The ratio $\left( d\Phi_\gamma^{Fermi} / dE \right) / \left( d\Phi_\gamma^{Theory} / dE \right)$ plotted as a function of the dimensionless energy ratio $x=E_\gamma /m_{S,V}$. This plot illustrates the case of a 65 GeV singlet, using the IGRB foreground model A for the fit. The three lines represent different values of the dark matter density $\rho_{\astrosun}$, demonstrating the effects of astrophysical uncertainties on the ratio.}
\end{figure}

It is necessary to emphasize that the mass region which is consistent with the Galactic center excess is not consistent with the current direct detection exclusion bounds as discussed in section 2. The LUX results rule out scalars below 100 GeV and vectors below around 200 GeV even with the most conservative estimates for the Higgs-nucleon interaction. The XENON100 results rule out vectors below around 130 GeV and scalars  below around 70 GeV. With this in mind, we can accept Higgs-portal singlets as an explanation for the Galactic center excess only if we disregard the bounds reported in \cite{Aprile:2012vw}\cite{PhysRevLett.109.181301}\cite{Faham:2014hza}. There has been work on simple modifications of the basic Higgs-portal scenario that would allow for both consistency with direct detecion results and an explanation of the Galactic center excess \cite{Wang:2014elb}.

\section{High mass region}

The lower mass region discussed in the previous section can potentially explain the Galactic center excess, but it is in mild tension with the diffuse spectrum observed by Fermi-LAT and much stronger tension with the nuclear recoil bounds reported by the LUX and XENON100 experiments. We explore a second mass region that is consistent with both the diffuse spectra and direct detection constraints. This mass region cannot explain the Galactic center excess through annihilation of Higgs-portal singlets, but it has other interesting features.

The mass region of 250 GeV to 1000 GeV is chosen for consistency with the Fermi IGRB spectrum. In Figure 8 the ratio of fluxes $\left( d\Phi_\gamma^{Fermi} / dE \right) / \left( d\Phi_\gamma^{Theory} / dE \right)$ is presented for selected masses in the range 250-1000 GeV. Both scalar and vector species are presented, for each of the three foreground models discussed in the reference \cite{Ackermann:2014usa}. As illustrated in Figure 8, this mass region is mostly acceptable for both the scalar and vector species with some variability due to the foreground model chosen. In choosing this mass range, a more conservative bound of a ratio of one has been used. In some regions, which have been marked, the predicted flux from Higgs-portal annihilations already exceeds the observed IGRB.

\begin{figure}
\begin{center}\includegraphics[scale=0.4]{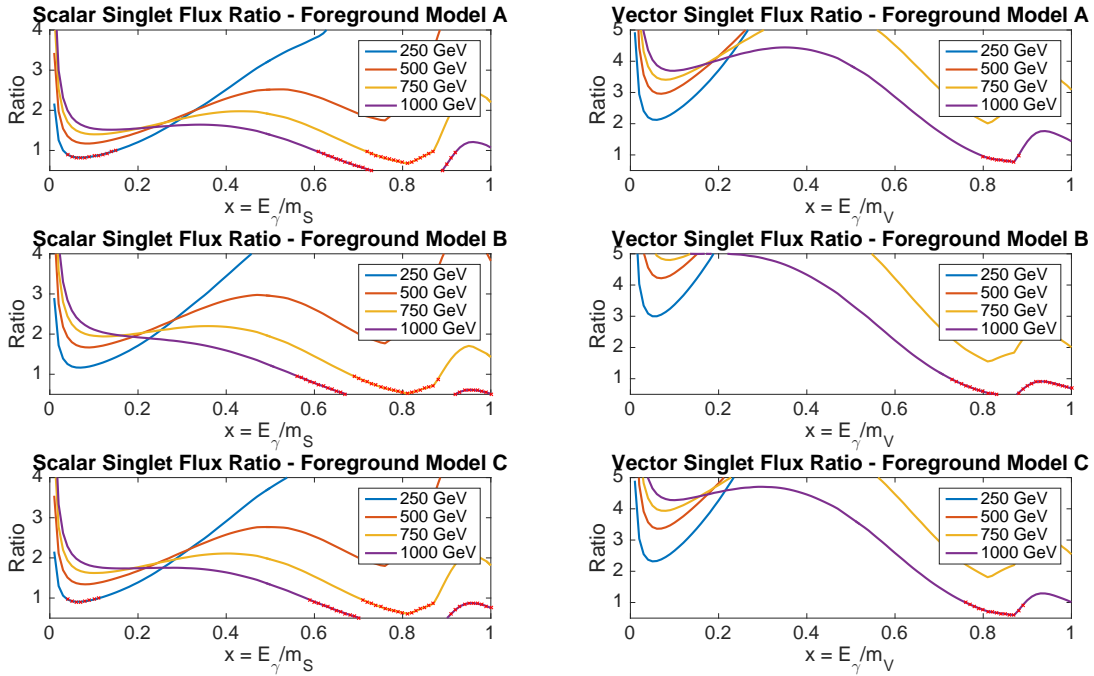}\end{center}
\caption{The ratio $\left( d\Phi_\gamma^{Fermi} / dE \right) / \left( d\Phi_\gamma^{Theory} / dE \right)$ plotted for each of the IGRB foreground models as a function of the dimensionless energy ratio $x=E_\gamma /m_{S,V}$. Masses are chosen to illustrate the regions where the theoretical flux is not consistent with the observed flux. Such regions are marked by the red `x's.}
\end{figure}

We note that 250 GeV Higgs-portal scalars are very close to producing an observable photon flux in the 15-25 GeV energy range. A vector singlet annihilation signal would be hidden deeper in the IGRB. Higher mass singlets are further from a ratio of one, but are still within an order of magnitude. It is possible that Fermi may be able to explore this region of the energy spectrum soon, allowing potential observation of a flux produced by the annihilation of scalar Higgs-portal singlets.

In the higher dark matter mass regions of hundreds to thousands of GeV, the observations of the Cherenkov Earth-based gamma ray telescopes become relevant. Due to high energy lower bounds, instruments of this type can say little about features that exist in the $E_\gamma < 10$ GeV range. However, they are capable of exploring the gamma ray spectrum up to 100s of TeV. This energy range can be used to impose bounds on the interactions of high mass dark matter particles. In the remainder of this section, we will use Cherenkov telescope observations of dwarf spheroidal galaxies to derive bounds on the Higgs-portal couplings $\eta^2_{S,V}$ and compare them against the thermally constrained couplings found in section 2. 

Dwarf spheroidal galaxies are small clusters of stars in the vicinity of the Milky Way that are believed to have a mass to luminosity ratio higher than any other object that has been observed within the local group of galaxis. Observations of stellar motion indicate they may contain thousands of times more dark matter than luminous matter. They are also fairly sterile objects; they have little star formation and few energetic events that can produce gamma rays. Additionally, they have very little gas that can produce gamma rays via Inverse Compton Scattering. These qualities make them promising targets for the observation of annihilation signals of cold dark matter. Indeed, the Earth-based Cherenkov gamma telescopes \cite{Aliu:2012ga}\cite{Abramowski:2014tra}\cite{Aleksic:2013xea} and Fermi-LAT \cite{Ackermann:2015zua} have all explored the possibility of these signals. None of them have yet reported a statistically significant excess, though independent analysis of Fermi data has revealed a potential excess in the dwarf galaxy Reticulum II \cite{PhysRevLett.115.081101}.

The dwarf galaxy Segue I is a satellite galaxy of the Milky Way, positioned at an angle well above the galactic disk. Stellar observations indicate that it is rich in dark matter. Both the VERITAS collaboration \cite{Aliu:2012ga} and the MAGIC collaboration \cite{Aleksic:2013xea} have recently reported event rates and an analysis of potential dark matter signals in Segue I. We discuss the bounds on the Higgs-portal singlet parameter space in this higher mass region that are provided by the observations of each of these collaborations in turn.

The MAGIC collaboration has provided bounds on dark matter annihilation cross sections through various channels which are reported in \cite{Ahnen:2016qkx}. These bounds are based off data from 158 hours of Segue I observations that was initially reported in \cite{Aleksic:2013xea}. The data was analyzed again in conjunction with Fermi data in \cite{Ahnen:2016qkx} using a modified statistical procedure. Full details of the statistical and astrophysical models are available in the reference \cite{Ahnen:2016qkx}.

Because our cross sections are proportional to the Higgs-portal coupling $\eta_{S,V}$ squared, we instead elect to extract the coupling dependence and present upper bounds on $\eta_{S,V}^2$. In Figure 9 we have provided 95\% confidence exclusion bounds on the Higgs-portal coupling in the mass range of 250-1000 GeV for both the scalar and vector models. The thermally constrained couplings are included for comparison. These bounds were found from the MAGIC data using the Higgs-portal annihilation spectra as discussed in section 3, rather than generic annihilation channels.

The number of events that we predict VERITAS will observe is given by

\begin{equation}
N_\gamma \left( E \ge E_{min} \right) = T_o \int_{E_{min}}^\infty dE \mathcal{A}_{eff}\left( E \right)\frac{d\Phi_\gamma}{dE}.\end{equation}

The differential photon flux $d\Phi_\gamma /dE$ is as described earlier, where we have taken the astrophysical factor to be as reported by the VERITAS collaboration \cite{Aliu:2012ga} $\bar{J}\left( \Delta\Omega\right) =7.7\times 10^{18}\:\mathrm{GeV}^2\:\mathrm{cm}^{-5}\:\mathrm{sr}$. The effective area function $\mathcal{A}_{eff}\left( E \right)$ is a function describing the instrument response to incoming gamma rays. It depends on the gamma ray energy and the zenith angle of the observation. Further details are available in the reference. We take values for the exposure time $T_o$ and the minimum energy $E_{min}$ from the reference \cite{Aliu:2012ga} as well.

As mentioned above, no statistically significant gamma ray excess has been observed in the Segue I dwarf galaxy. This has led the VERITAS collaboration to report bounds on the annihilation cross section $\langle\sigma v\rangle$ for a generic WIMP. These bounds are provided in a model-independent fashion, assuming a generic photon production spectrum and specific annihilation channels. We perform a similar analysis for Higgs-portal models in the heavier mass range described at the beginning of this section.

The above relation (5.1) can be inverted and by inserting the reported 95\% confidence event rates, a bound can be placed on the annihilation cross section. These bounds are plotted as a function of mass in comparison to the MAGIC bounds and the thermally constrained couplings in Figure 9.

\begin{figure}
\begin{center}\includegraphics[scale=0.4]{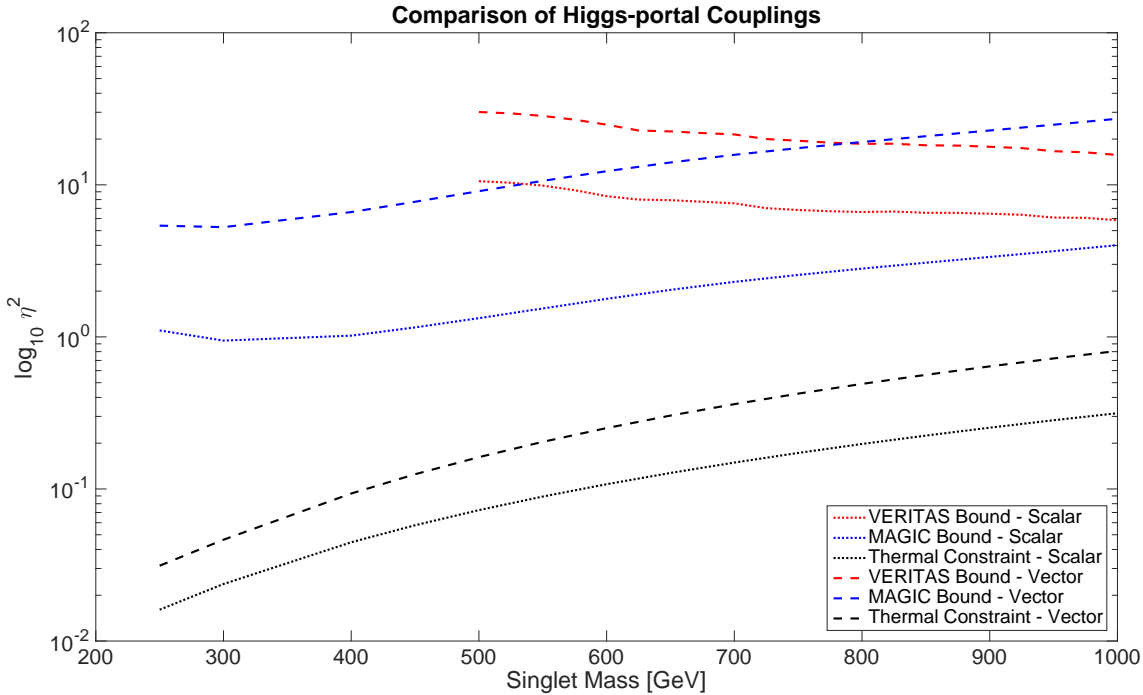}\end{center}
\caption{Bounds from Segue I on Higgs-portal coupling derived from VERITAS (red) and MAGIC (blue) observations for the scalar and vector singlet models as a function of singlet mass. Thermally constrained couplings (black) over the same mass range are included for comparison.}
\end{figure}

The VERITAS telescope is largely insensitive to energies below 300 GeV \cite{Aliu:2012ga}. Taking this into account, we neglect anything below 500 GeV for the purposes of this comparison. We note that these bounds are consistent with the bounds on the annihilation cross section of a generic WIMP as calculated by the VERITAS collaboration \cite{Aliu:2012ga}.

The bounds from MAGIC observations are stronger than those from VERITAS for most of the mass range that was explored.  The thermally constrained couplings are clearly an order of magnitude lower than the Segue I observational bounds from both collaborations, so no mass exclusions can be made. However, bounds in the lower mass region for the scalar model are of order unity, within the perturbative regime. These are the strongest bounds that have been placed on the Higgs-portal coupling from indirect detection. With more observation time, these bounds can be lowered further to the point where exclusions of the parameter space can be made.

\section{Conclusions}

This article has discussed potential gamma ray signals of annihilating scalar and vector Higgs-portal dark matter in the mass range $m_{S,V} \ge 63$ GeV. We have constrained the Higgs-portal coupling by assuming that Higgs-portal singlets make up all of the currently observed cold dark matter. Singlets with masses below $m_h/2$ are excluded by strong bounds on invisible Higgs boson decays as reported by ATLAS and CMS.

We have found that both the scalar and vector species with a mass of $~65$ GeV are consistent with the Galactic center excess. This mass is, however, not consistent with the bounds on the diffuse annihilation signal as obtained from the observations of the IGRB by Fermi-LAT. It also is inconsistent with current bounds on the nuclear recoils cross sections as reported by the XENON100 and LUX experiments.

Singlets in the mass range fo 250-1000 GeV are consistent with both nuclear recoil cross section bounds and observations of the IGRB, and might be observable in the near future. Bounds on the Higgs-portal couplings from observations of the dwarf galaxy Segue I by the MAGIC and VERITAS collaborations are calculated and compared to the thermally constrained couplings.

\subsection*{Acknowledgments}

The authors would like to thank Ben Zitzer and David Hanna for their assistance with the portion of this article that deals with VERITAS observations of Segue I. We would also like to thank the MAGIC collaboration and Javier Rico in particular for providing the bounds on the annihilation cross sections that were used in section 5. This work was supported in part by the Natural Sciences and Engineering Research Council of Canada (NSERC) Discovery Grant program.

\section*{Appendix A: Explicit Cross Sections}

The relativistic annihilation cross sections to leading order are as reported in \cite{Sage2015}. We reproduce them here:

\begin{equation}
\sigma_{SS\rightarrow hh}=\frac{\eta_S^2}{32\pi}\frac{\sqrt{k^2+m_S^2-m_h^2}}{k\left(k^2+m_S^2\right)}\frac{\left(2k^2+m_S^2+m_h^2\right)^2}{\left(4k^2+4m_S^2-m_h^2\right)^2+m_h^2\Gamma_h^2},\end{equation}

\begin{equation}
\sigma_{SS\rightarrow f\bar{f}}=\frac{\eta_S^2 N_C}{8\pi}\frac{\left(k^2+m_S^2-m_f^2\right)^{3/2}}{k\left(k^2+m_S^2\right)}\frac{m_f^2}{\left(4k^2+4m_S^2-m_h^2\right)^2+m_h^2\Gamma_h^2},\end{equation}

\begin{equation}
\sigma_{SS\rightarrow ZZ,WW}=\frac{\eta_S^2}{16\pi (1+\delta_Z)}\frac{\sqrt{k^2+m_S^2-m_{Z,W}^2}}{k\left(k^2+m_S^2\right)}\frac{\left[2m_{Z,W}^4+\left(m_{Z,W}^2-2k^2-2m_S^2\right)^2\right]}{\left(4k^2+4m_S^2-m_h^2\right)^2+m_h^2\Gamma_h^2},\end{equation}

\begin{equation}
\sigma_{VV\rightarrow hh}=\frac{\eta_V^2}{288\pi}\frac{\sqrt{k^2+m_V^2-m_h^2}}{m_V^4 k\left(k^2+m_V^2\right)}\left( \left(2k^2+m_V^2\right)^2+2m_V^4\right)\left(\frac{2k^2+m_V^2+m_h^2}{4k^2+4m_V^2-m_h^2}\right)^2,\end{equation}

\begin{equation}
\sigma_{VV\rightarrow f\bar{f}}=\frac{\eta_V^2 N_C m_f^2}{72\pi m_V^4}\frac{\left(k^2+m_V^2-m_f^2\right)^{3/2}}{k\left(k^2+m_V^2\right)}\frac{\left(2k^2+m_V^2\right)^2+2m_V^4}{\left(4k^2+4m_V^2-m_h^2\right)^2+m_h^2\Gamma_h^2},\end{equation}

\[
\sigma_{VV\rightarrow ZZ,WW}=\frac{\eta_V^2 \sqrt{k^2+m_V^2-m_{Z,W}^2}}{144\pi m_V^4(1+\delta_Z)}\frac{\left[\left(2k^2+m_V^2\right)^2+2m_V^4\right]}{k\left(k^2+m_V^2\right)}\]
\begin{equation}
\times\frac{\left[2m_{Z,W}^4+\left(2k^2+2m_V^2-m_{Z,W}^2\right)^2\right]}{\left(4k^2+4m_V^2-m_h^2\right)^2+m_h^2\Gamma_h^2}.\end{equation}

In the above, $k$ is the magnitude of the three-momentum of the initial state annihilating particles in the center of momentum frame. Masses are as labelled, and the fermion cross section is scaled by the number of colors, where $N_C=1$ for leptons and $N_C=3$ for quarks. In annihilation to electroweak vector boson cross sections $\delta_Z=1$ for annihilation into the $ZZ$ final state and $\delta_Z=0$ for annihilation into the $W^+W^-$ final state. We have employed the Breit-Wigner prescription for the propagator of an unstable particle, leading to a term in the denominator containing the Higgs boson decay width $\Gamma_h$. 

The following are expressions for the annihilation of scalar and vector Higgs-portal singlets to photons, which occurs at leading order through loop effects. These quantities have appeared in the literature before \cite{Duerr:2015aka}, calculated from the full loop diagrams. We present here an alternate version calculated using effective Higgs-photon interactions as detailed in \cite{Vainshtein1979}. This method allows for the estimation of the cross sections for direct annihilation to photons without the necessity of dealing with loop integrations. Similar expressions exist for direct annihilation to gluons.

\begin{equation}
\sigma_{SS\rightarrow\gamma\gamma}=\frac{g_{h\gamma\gamma}^2 v_h^2 \eta_S^2 \left(k^2+m_S^2\right)^{3/2}}{1024\pi k \left[ \left( 4k^2+4m_S^2-m_h^2\right)^2+m_h^2\Gamma_h^2\right]},\end{equation}

\begin{equation}
\sigma_{VV\rightarrow\gamma\gamma}=\frac{g_{h\gamma\gamma}^2 v_h^2 \eta_V^2}{128\pi k}\frac{\left(k^2+m_V^2\right)^{3/2}}{\left[\left(4k^2+4m_V^2-m_h^2\right)^2+m_h^2\Gamma_h^2\right]}\left( 2+\frac{\left(m_V^2+2k^2\right)^2}{m_V^4}\right).\end{equation}

\bibliographystyle{h-physrev}
\bibliography{HPGAMREFS}

\end{document}